\documentclass[twoside,slac_one]{revtex4}
\usepackage{graphicx}
\usepackage{fancyhdr}
\usepackage{amsmath} 
\usepackage{bm}
\usepackage{amsxtra}
\usepackage{amssymb}
\usepackage{amsthm}
\usepackage{latexsym}
\usepackage{lscape}
\usepackage{subfigure}

\def\cm{\hbox{$\;\hbox{\rm cm}$}}
\def\TeV{\hbox{$\;\hbox{\rm TeV}$}}
\newcommand{\GeVcc}{\ensuremath{\,\mathrm{Ge\kern -0.1em V\!/c}^2}}
\newcommand{\GeVc}{\ensuremath{\mathrm{Ge\kern -0.1em V}\!/c}}
\def\GeV{\hbox{$\;\hbox{\rm GeV}$}}
\newcommand{\ifb}{\ensuremath{\mathrm{fb^{-1}}}}

\begin{document}

\title{A search for the Higgs Boson in H$\rightarrow$ZZ$\rightarrow$4l mode}
\author{M. Pelliccioni on behalf of CMS Collaboration}
\affiliation{Istituto Nazionale di Fisica Nucleare, Torino, IT}

\begin{abstract}
A search for a Higgs boson in the decay channel ${\rm H} \rightarrow {\rm ZZ}^{(*)}$ with each Z boson decaying to an 
electron or muon pair is presented using pp collisions from the LHC at $\sqrt s$ = 7 TeV.
The data analyzed correspond to an integrated luminosity of $1.13 \pm 0.07$~fb$^{-1}$ recorded by the 
CMS detector in 2010 and 2011.
The search covers Higgs boson mass ($m_{\rm H}$) hypotheses of $110 < m_{\rm H} < 600 \GeVcc$.
Fifteen events are observed, while $14.4 \pm 0.6$ events are expected from standard model background processes. 
Upper limits at 95\% CL on the cross section$\times$branching ratio 
for a Higgs boson with  standard model-like decays exclude cross sections from about one to two times the expected 
standard model cross section for masses in the range $180 < m_{\rm H} < 420 \GeVcc$. 
Reinterpreted in the context of the standard model with four fermion families a Higgs boson with a mass in the range 
138 -162  $\GeVcc$  and 178-502 $\GeVcc$  is excluded at 95\% CL.
\end{abstract}

\maketitle

\thispagestyle{fancy}


\section{Introduction}\label{sec:Introduction}

The standard model (SM) of electroweak interactions~\cite{StandardModel67} 
predicts the existence of a scalar boson, the Higgs boson,  associated to the spontaneous electroweak symmetry 
breaking~\cite{Higgs:1964}.
The mass $m_{\rm H}$ of this scalar boson is a free parameter of the theory.
The inclusive production of SM Higgs bosons followed by the decay H$\rightarrow$ZZ$^{(*)}$ is expected to be a 
main discovery channel at the CERN LHC pp collider.

The direct searches for the SM Higgs boson at the LEP $\rm e^+e^-$ collider have lead to a lower mass  bound of 
$m_{\rm H} > 114.4$~\GeVcc (95\% CL)~\cite{Barate:2003sz}.
The direct searches by the D0 and CDF experiments at the Tevatron exclude the mass range
$158 < m_{\rm H} < 173$~\GeVcc (95\% CL)~\cite{Aaltonen:2011gs}.
Indirect constraints from precision measurements favour the mass range $m_{\rm H} < 185$~\GeVcc (95\% CL)~\cite{:2010zz}.

A search for a SM Higgs boson in the four-lepton decay channel,
H$\rightarrow$ZZ$^{(*)} \rightarrow  \ell^{\pm}\ell^{\mp}\ell^{'\pm}\ell^{'\mp}$
with $\ell, \ell' = e$ or $\mu$, in short H$\rightarrow 4\ell$, is presented.
The analysis is designed for a Higgs boson mass in the range $110 < m_{\rm H} < 600 ~\GeVcc$
and uses data collected by the CMS experiment during 2010 and 2011 at the LHC collider,  with pp collisions 
at $\sqrt s$ = 7~TeV.
The sample of data corresponds to an integrated luminosity of  $ {\cal L} = 1.13  \pm 0.07 \, \ifb $.

The sample of events with four reconstructed leptons contains an irreducible contribution 
from $ZZ^{(*)}$ production via $q\bar{q}$ and $gg$ fusion processes. 
Potential reducible background contributions are from ${\rm Z}b\bar{b}$ and 
$t\bar{t} \rightarrow {\rm W^+} b {\rm W^-} {\bar b}$, with the W undergoing a leptonic decay, 
where the final states contain two isolated leptons.
Reconstructed $4\ell$ events can also arise from instrumental background such as Z+jets or WZ + jet(s)
or from the production of multiple jets in QCD hard interactions, where jets are misidentified as leptons. 

A detailed description of the CMS detector can be found elsewhere~\cite{:2008zzk}. 

The trigger used in this analysis evolved in response to the increasing instantaneous luminosity.
The presence of a pair of electrons with $E_{T,1}^e > 17$ and $E_{T,2}^e > 8 \, \GeV$, or a pair of muons
with $p_{T,1}^{\mu} > 13$ and $p_{T,2}^{\mu} > 8 \, \GeVc$, is required for a major fraction of the data sample, 
collected with instantaneous luminosities up to $\simeq 1.3 \times 10^{33} \cm^{-2} s^{-1}$. 
The trigger is fully efficient for the $4e$, $4\mu$ and $2e2\mu$ channels.

Monte Carlo (MC) samples for the SM Higgs boson signal and a variety of electroweak 
and QCD-induced SM background processes have been used for the optimization of the event selection prior to the analysis.
They are also used for comparisons to the measurements, and the evaluation of acceptance corrections and systematic uncertainties.
All production cross sections are re-weighted at least to next-to-leading-order (NLO), or beyond 
where possible. The general multi-purpose MC event generator {\tt PYTHIA}~\cite{Sjostrand:2006za} 
is used in conjunction with other event generators, for the showering, hadronization, decays and to add the 
underlying pp event. 
The events are processed with a detailed simulation of the CMS detector based on {\tt GEANT4}~\cite{GEANT}. 
The Higgs boson samples are generated with {\tt POWHEG}~\cite{powheg} 
which incorporates NLO gluon fusion and weak-boson fusion.
The events are re-weighted according to the most recent calculations~\cite{LHCHiggsCrossSectionWorkingGroup:2011ti}
of the total cross section $\sigma(pp\rightarrow {\rm H})$  which comprises gluon and weak-boson fusion contributions.
The total cross section is scaled by the ${\cal B}({\rm H}\rightarrow 4\ell)$ 
\cite{LHCHiggsCrossSectionWorkingGroup:2011ti,hdecay2,Actis:2008ts}.
Interference in the $4e$ or $4\mu$ channels is taken into account using the 
{\tt Prophecy4f}~\cite{LHCHiggsCrossSectionWorkingGroup:2011ti}
generator tool for precision calculations.
Di-boson production (WW, WZ, ZZ, Z$\gamma$) is generated at leading order (LO) with {\tt PYTHIA}.
Interference effects involving the Higgs boson are neglected.
An uncertainty of $\pm30\,\%$ is attributed the gluon-induced contribution.
An inclusive Z+jets sample has been generated with {\tt MadGraph}~\cite{Alwall:2007st},
considering all (five) quark flavours for the initial state parton density functions (PDFs),
and gluons, light ($d,u,s$) and heavy ($c,b$) quarks for the jets in the final state.
A K-factor is used to correct to the NLO cross section.
A top pair production sample is generated at NLO with {\tt POWHEG}.

\section{Event Selection}\label{sec:Selection}

The selection acts on loosely isolated (see below) lepton candidates, i.e. electrons within the 
geometrical acceptance of $|\eta^e| < 2.5$ and $p_T^e > 7~\GeVc$ and muons satisfying 
$|\eta^\mu| < 2.4$ and $p_T^\mu > 5~\GeVc$. We require:
\begin{enumerate}
\item {\it First Z:}
         a pair of lepton candidates of opposite charge and matching flavour ($e^+ e^-$, \, $\mu^+\mu^-$) 
         satisfying $m_{1,2} > 60~\GeVcc$, $p_{T,1} > 20~\GeVc$ and $p_{T,2} > 10~\GeVc$; the pair
         with reconstructed mass closest to the nominal Z boson mass is retained and denoted ${\rm Z_1}$.         
\item {\it Choice of the ``best $4\ell$'':} retain a second lepton pair, denoted ${\rm Z_2}$, 
         among all the remaining $\ell^+ \ell^-$ combinations with $m_{\rm Z_2} > 12\GeVcc$ and such that 
         the reconstructed four-lepton invariant mass satisfies  $m_{4\ell} > 100~\GeVcc$.
         For the $4e$ and $4\mu$ final states, at least three of the four combinations of opposite sign 
         pairs must satisfy $m_{\ell\ell} > 12\GeVcc$.
         If more than one ${\rm Z_2}$ combination satisfies all the criteria, 
         the one built from leptons of highest $p_T$ is chosen.
\item {\it Relative isolation for selected leptons:}  \, for any combination of two leptons $i$ and $j$, 
          irrespective of flavour or charge,  
          the sum of the combined relative isolation $R_{iso,j} +  R_{iso,i} < 0.35$.
\item {\it Impact parameter for selected leptons:} \,   the significance of the impact
          parameter to the event vertex, ${\rm SIP_{3D}}$,  is required to satisfy  $| {\rm SIP_{3D}} = \frac{\rm IP}{\sigma_{\rm IP}}  | < 4$ 
          for each lepton, where ${\rm IP}$  is the lepton impact parameter in
          three dimensions at the point of closest approach with respect to the
          primary interaction vertex, and  $\sigma_{\rm IP}$ the associated uncertainty. 
\item  {\it ZZ$^{(*)}$ kinematics:} $60 < m_{\rm Z_1} < 120~\GeVcc$ and $m_{\rm Z}^{min} < m_{\rm Z_2} < 120~\GeVcc$,
            where $m_{\rm Z}^{min}$ is defined below.           
\end{enumerate}                  

Requiring all selection criteria with  $m_{\rm Z}^{min} \equiv 20~\GeVcc$ defines the {\bf baseline selection}
which is used in subsequent analysis, independent of the $m_{\rm H}$ hypothesis.
Requiring all selection criteria with $m_{\rm Z}^{min} \equiv 60~\GeVcc$ defines the {\bf high-mass selection}
which is used for the measurement of the ZZ cross section and for the Higgs boson searches at high mass
($m_{\rm H} > 2 \times m_{\rm Z}$).

The efficiencies for reconstruction, identification and selection for electrons and muons is
measured with data by using a ``tag-and-probe'' technique based on the sample of 
inclusive single Z production events. 
The measurements have been performed in several ranges in $  |\eta| $ matching 
domains with uniform detector and reconstruction performances for the tracker and 
calorimeters, and in $p_T$ ranges from $7 \GeVc$ to $100 \GeVc$.

The reconstructed leptons  must also satisfy a very loose relative "track-only"  isolation 
$ R_{\rm Iso}^{\rm track} < 0.7$ where $ R_{\rm Iso}^{\rm track} = (1/ p_T^{\ell}) \times A_{\rm Iso}^{\rm track}$
and $A_{\rm Iso}^{\rm track} = \sum_{i}  p_{T,{\rm track}}^i  $.
The sum runs over the tracks $i$ within a cone of radius $\Delta R = \sqrt{\Delta\eta^{2} + \Delta\phi^{2}} < 0.3$
with axis along the lepton candidate direction, excluding a central inner "veto" region.
For the leptons chosen to form the $4\ell$ system in event candidates, the combination of the 
tracker, ECAL and HCAL information is then used for isolation.
The combined lepton $R_{iso}$ is calculated as 
$R_{iso} = (1/ p_T^{\ell}) \times \left( A_{\rm Iso}^{\rm track} +  A_{\rm Iso}^{\rm ECAL} + A_{\rm Iso}^{\rm HCAL} \right)$.
The  $A_{\rm Iso}^{\rm ECAL}$ and $A_{\rm Iso}^{\rm HCAL}$ are sums over the $E_T$ from energy deposits in cells 
of the ECAL and HCAL respectively, and with geometrical centroids situated within a cone of radius  
$\Delta R = \sqrt{\Delta\eta^{2} + \Delta\phi^{2}} < 0.3$. 
The measurements are performed for different $  |\eta| $ ranges covering the geometrical
acceptance and $p_T$ ranges up to $100 \GeVc$.
The combined isolation efficiencies measured with data using tag-and-probe
is found to be above 99\% everywhere for muons and between 94\% and 99\% for electrons.

The overall signal detection efficiency for a $4\ell$ system within the geometrical acceptance is evaluated 
from MC to be rising from about 42\%~/~72\%~/~54\% at $m_{\rm H} = 190 \GeVcc$ 
to about 59\% / 82\% / 71\% at $m_{\rm H} = 400 \GeVcc$ for the  $4e$~/~$4\mu$~/~$2e2\mu$ channels.
A fit of the signal mass distribution as obtained from the MC shows a resolution for a Higgs boson 
mass hypothesis of $150 \GeVcc$ in the $4e$ ($4\mu$, $2e2\mu$) 
of $2.7 (1.6, 2.1) \pm 0.1 \GeVcc$.
Just above the $2 \times m_{\rm Z}$ threshold, for $m_{\rm H} = 190 \GeVcc$, the resolution
for $4e$ ($4\mu$, $2e2\mu$) is $3.5 (2.5, 2.8) \pm 0.1 \GeVcc$.

\section{Measurement and Control of the Background}\label{sec:Background}

The small number of events observed precludes a precise evaluation of the background using 
mass sidebands.
We therefore rely on other methods that use data to estimate the backgrounds
and the associated systematic uncertainties.
We choose a wide background control region outside  the signal 
phase space which is populated by relaxing some selection criteria.

\subsection{ZZ$^{(*)}$ Continuum}\label{sec:Continuum}

Two different methods have been used to determine $N^{\rm ZZ}_{\rm expect}$ for the ZZ$^{(*)}$
di-boson continuum,  a normalization to the measured Z rate and an estimate from MC simulation.  
The former is our reference method for the present low integrated luminosities. 
A direct measurement from sidebands, e.g. excluding the signal region around a given Higgs boson
mass hypothesis, is affected by large statistical uncertainties and will become useful only at
higher integrated luminosities.

The method of normalization to the measured Z rate relies on the measurement of inclusive single Z production which is used to predict the total 
ZZ rate within the acceptance defined by this analysis, making use of the ratio of the theoretical
cross sections for Z and ZZ production, and of the ratio of the reconstruction and selection
efficiencies for the $2\ell$ and  $4\ell$  final states. 
The ZZ$^{(*)} \rightarrow 4 \ell$ control region is defined here by the observed single Z inclusive
rate for ${\rm Z} \rightarrow \ell \ell$. 
The ratio of acceptance between the control and the signal region is then given by combining the 
theoretical  cross sections and the selection efficiencies as obtained from MC simulation:
\begin{equation}
R^{\sigma}_{\rm theory} \times R^{\epsilon}_{\rm MC} 
=
\frac{\sigma^{q\overline{q} \rightarrow {\rm ZZ} \rightarrow 4\ell}_{\rm NLO} 
+ \sigma^{gg \rightarrow {\rm ZZ} \rightarrow 4\ell}_{\rm LO}}{\sigma^{pp \rightarrow {\rm Z} \rightarrow 2\ell}_{\rm NNLO}} 
\times
\frac{\epsilon^{ {\rm ZZ} \rightarrow 4\ell}_{\rm MC}}{\epsilon^{ {\rm Z} \rightarrow 2\ell}_{\rm MC}} \,\,.
\label{eq_ratio} 
\end{equation}

The cross section for ZZ$^{(*)}$ production at NLO through qq annihilation and gg fusion are calculated with {\tt MCFM},
while the cross section for Z at NNLO is calculated with FEWZ 2.0. 
The theoretical uncertainties on the ratio are computed varying for each $4\ell$ and $2\ell$ final states 
both the QCD renormalization and factorization scales ($m_{\rm Z}$, $m_{\rm Z}/2$, 2$m_{\rm Z}$).

The efficiency $\epsilon_{\rm MC}$ is defined as the ratio between events passing all the selection criteria and events 
with the cut at generator level ($m_{ll} > 12$ GeV/$c^2$).
At the reconstruction level, the ZZ$^{(*)}$ events must fulfill all the event selection criteria.
The selection of Z events follows from the first step of the selection.

In a complementary method the estimation of the number of ZZ events in any given mass 
range $\left[ m_1, m_2 \right]$ is obtained directly from the absolute rate predicted by the MC 
model simulation.

The number of events and relative uncertainties from $ZZ^{(*)}\rightarrow 4\ell$ 
predicted by normalization to the measured Z rate and 
by the MC model simulation for an integrated luminosity of 1.13$ \,\ifb$  in the signal region in a mass range from 100 to
600~\GeVcc with the baseline and high-mass selections are given in Table~\ref{tab:ZZBack}.

\begin{table}[htb]
\begin{center}
\caption{Number of ZZ background events and relative uncertainties in the signal region in a mass range from 100 to
600~\GeVcc, 
estimated from normalization to the measured Z rate  and 
from Monte Carlo simulation.}
\label{tab:ZZBack}
\begin{tabular}{c|c|cc}
\hline
& channel & \text{ Normalization to Z  rate} & \text{ MC model simulation } 
\\
\hline
baseline & $N^{{\rm ZZ} \rightarrow 4e}$  & 2.76 $\pm$ 0.18 & 2.77 $\pm$ 0.26 \\
& $N^{{\rm ZZ} \rightarrow 4\mu}$ & 4.10 $\pm$ 0.27 & 4.24 $\pm$ 0.39 \\
& $N^{{\rm ZZ} \rightarrow 2e2\mu}$ &  6.72 $\pm$ 0.45 & 6.85 $\pm$ 0.63 \\
\hline
high-mass & $N^{{\rm ZZ} \rightarrow 4e}$  &  2.50 $\pm$ 0.17 & 2.52 $\pm$ 0.23 \\
& $N^{{\rm ZZ} \rightarrow 4\mu}$ & 3.55 $\pm$ 0.23 & 3.66 $\pm$ 0.33 \\
& $N^{{\rm ZZ} \rightarrow 2e2\mu}$ &  6.10 $\pm$ 0.40 & 6.22 $\pm$ 0.58 \\
\end{tabular}
\end{center}
\end{table}

\subsection{Reducible and Instrumental Backgrounds}\label{sec:ReducibleBackground}

The Z$b\bar{b}/c\bar{c}$ and $t\bar{t}$ control region is defined taking a set of events with
a pair of identified leptons with opposite charge and matching flavour ($e^{+}e^{-}, \mu^{+}\mu^{-}$) that
form a good Z$_1$, as requested for the signal selection.
For the other pair of leptons the flavour, charge, and isolation requirements are removed. 
Finally the SIP$_{\rm 3D}$ impact parameter cuts are reversed on the two leptons by 
requiring $| {\rm SIP_{3D}} | > 5$.
Inverting the SIP$_{\rm 3D}$ selection cuts for two leptons ensures a negligible Z+jets contribution 
in the four-lepton background control region.

To extract the number of  $t\bar{t}$  and Z$b\bar{b}/c\bar{c}$ events in the four-lepton signal region,
we exploit the knowledge and distinct features of the ${\rm SIP_{3D}}$ distribution.
The $ {\rm SIP_{3D}} $ distributions for the ${\rm Z_2}$ leptons of the $t\bar{t}$  and Z$b\bar{b}/c\bar{c}$ backgrounds 
are uniform and of similar shapes. 
This is in sharp contrast with the expected $ {\rm SIP_{3D}} $ distribution for the 
signal which is concentrated at low $ {\rm SIP_{3D}} $ and steeply falling with increasing $ {\rm SIP_{3D}} $. 

In order to control the Z+jets background, a four-lepton background control region is obtained by relaxing the 
cuts on isolation and identification requirements for two additional leptons. 

Normalized to the integrated luminosity, the number of events from $t\bar{t}$, Z$b\bar{b}/c\bar{c}$ and Z+jets expected 
in the signal region in a mass range from $m_1 = 100~\GeVcc$ to $m_2 = 600~\GeVcc$ is given in  Table~\ref{tab:IrrBack}.

\begin{table}[htb]
\begin{center}
\caption{Number of background events for baseline and high-mass event selections 
                in the signal region in a $m_4\ell$ range from $100$ to $600~\GeVcc$, estimated from 
                data.}
\label{tab:IrrBack}
\begin{tabular}{c|c|c}
& \text{ baseline  } & \text{ high-mass } \\
\hline
$N^{{\rm Z}b\bar{b}/c\bar{c}, t\bar{t} \to 4e}$ & 0.01 $\pm$ 0.02 & - 
\\
$N^{{\rm Z}b\bar{b}/c\bar{c}, t\bar{t}\to 4\mu}$ & 0.01 $\pm$ 0.01& -
\\
$N^{{\rm Z}b\bar{b}/c\bar{c}, t\bar{t} \to 2e2\mu}$ & 0.02 $\pm$ 0.02 & -
\\
\hline
$N^{{\rm Z+jets} \to 4e}$ &  0.37 $\pm$ 0.07 & 0.14$\pm$ 0.06     \\ 
$N^{{\rm Z+jets}\to 4\mu}$ & 0.06 $\pm$ 0.01 & 0.004 $\pm$ 0.004  \\ 
$N^{{\rm Z+jets}\to 2e2\mu}$ &  0.39 $\pm$ 0.07 & 0.15 $\pm$ 0.06 \\ 
\end{tabular}
\end{center}
\end{table}

\section{Systematic uncertainties}\label{sec:Systematics}

The main sources of systematic uncertainties on the expected yields are summarized in Table~\ref{Tab:systematics}.
Systematic uncertainties have been evaluated from data for the trigger efficiency 
as well as for effects from individual lepton reconstruction, identification and isolation efficiencies, and from
energy-momentum calibration.  Additional systematics
come from the limited statistics in the background control regions which propagate to the
background evaluation in the signal region.
All major background sources are derived from control regions, and the comparison of
the data with the background expectation in the signal region is independent of the 
uncertainty on the LHC integrated luminosity for the data sample.
This uncertainty enters in the calculation of a cross section limit through the normalization
of the signal. 
Also given for completeness in Table~\ref{Tab:systematics} is an evaluation of the 
systematics uncertainties on the Higgs boson cross section and branching ratios.

\begin{table}[htb]
\begin{center}
\caption{Summary of the magnitude of systematic uncertainties in percent.}
\label{Tab:systematics} 
\begin{tabular}{c|c}
\hline
Luminosity                                        & 6   \\
Trigger efficiency                            & 1.5  \\ 
Higgs cross section                       & 17-20   \\
Higgs B.R.                                       & 2   \\
Lepton reco/ID eff.                         & 2-3  \\
Lepton isolation eff.                      & 2   \\ 
Electron energy scale                     &  3      \\ 
\hline
\end{tabular}
\end{center}
\end{table}

\section{Results}\label{sec:Results}

The reconstructed four-lepton invariant mass distributions obtained in the $4e$, $4\mu$, and  $2e2\mu$ channels
with the baseline selection are shown in Fig.~\ref{fig:Mass4lBaseline}, 
and compared to expectations from  the SM backgrounds. 
Three (six, six) event candidates are observed in $4e$ ($4\mu$, $2e2\mu$) final states,
satisfying the baseline selection.

\begin{figure}[!htb]
\begin{center}
\begin{tabular}{cc}
\subfigure[]{\includegraphics[width=0.5\textwidth]{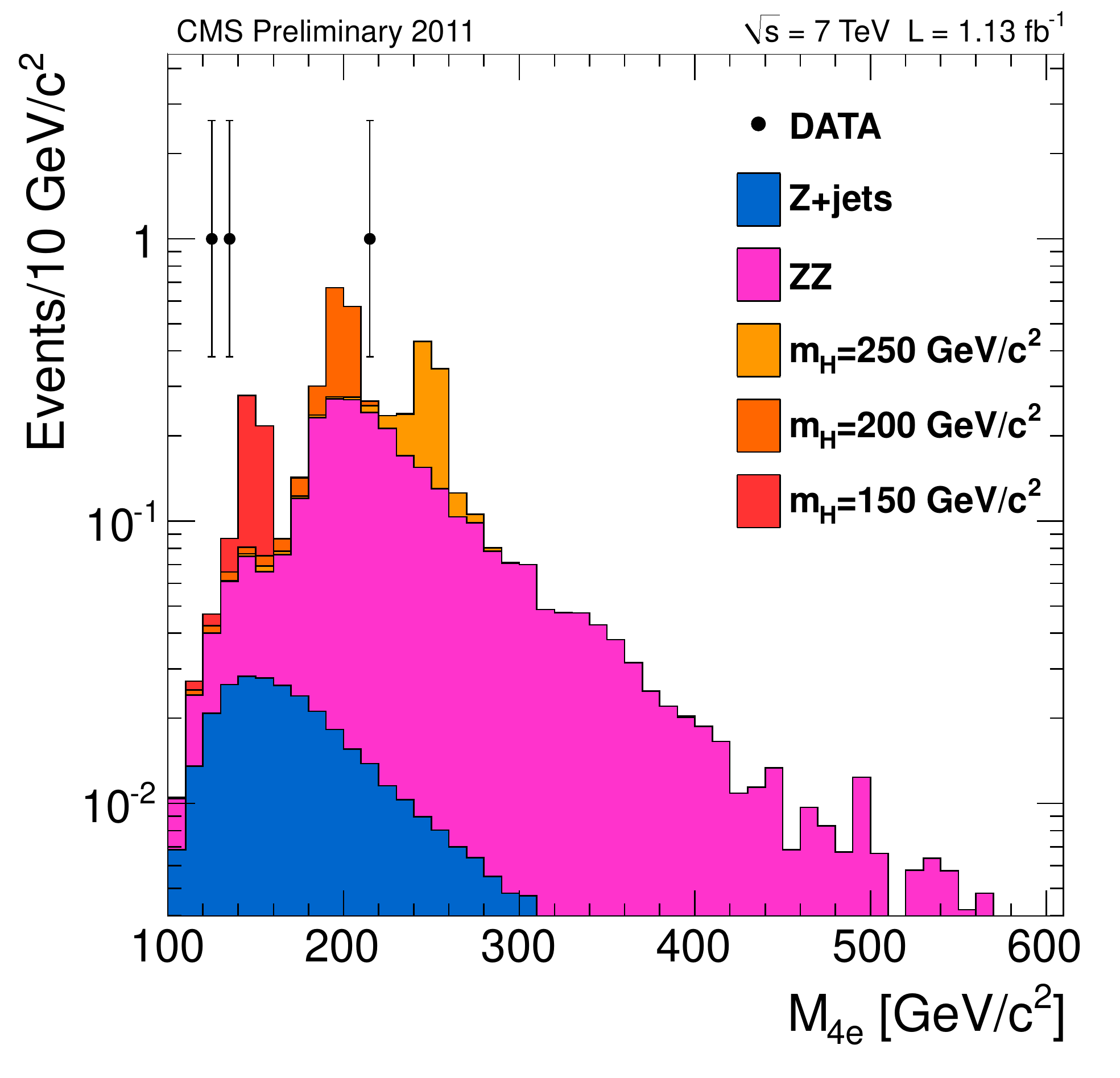}} 
\subfigure[]{\includegraphics[width=0.5\textwidth]{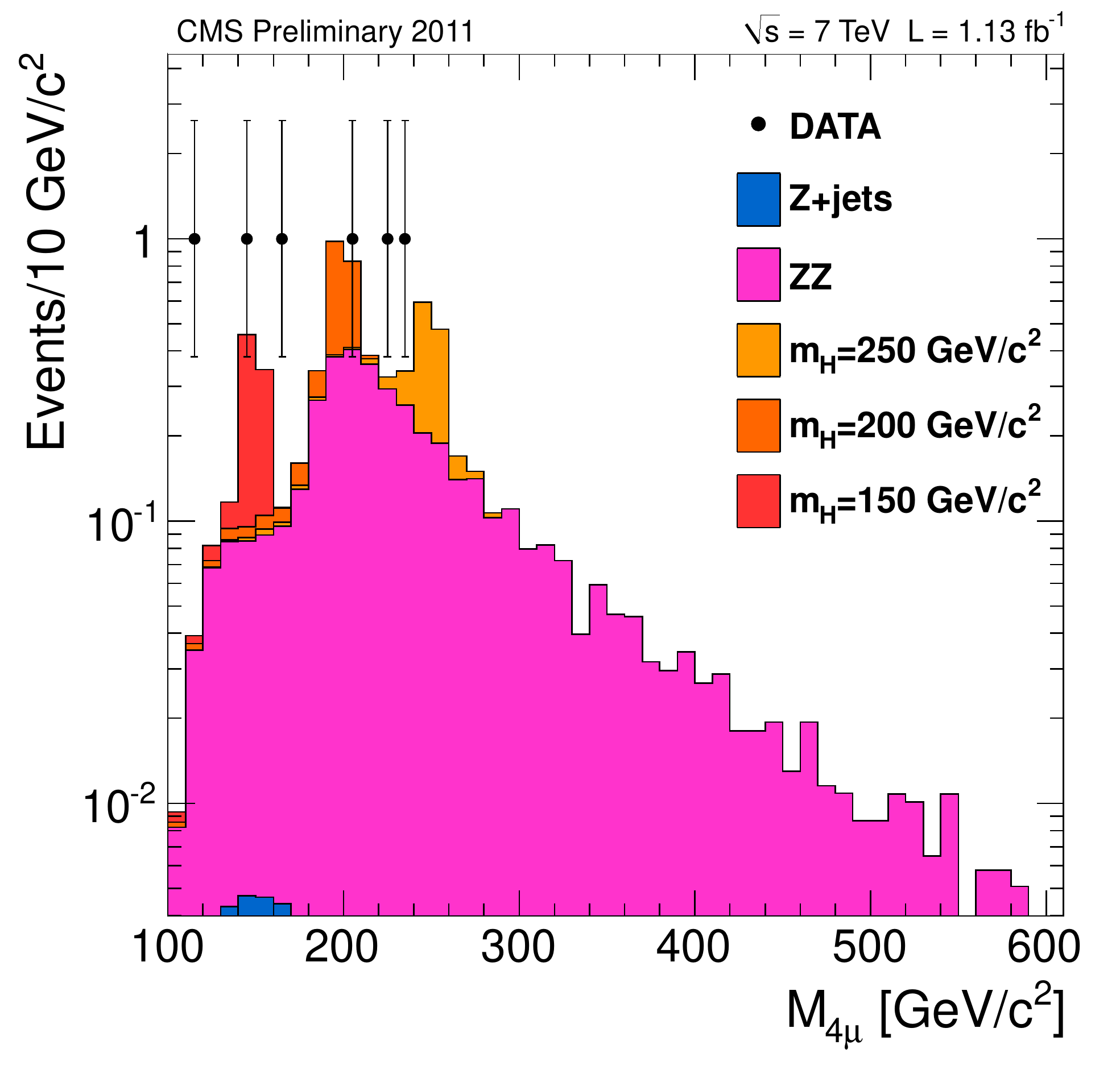}} \\
\subfigure[]{\includegraphics[width=0.5\textwidth]{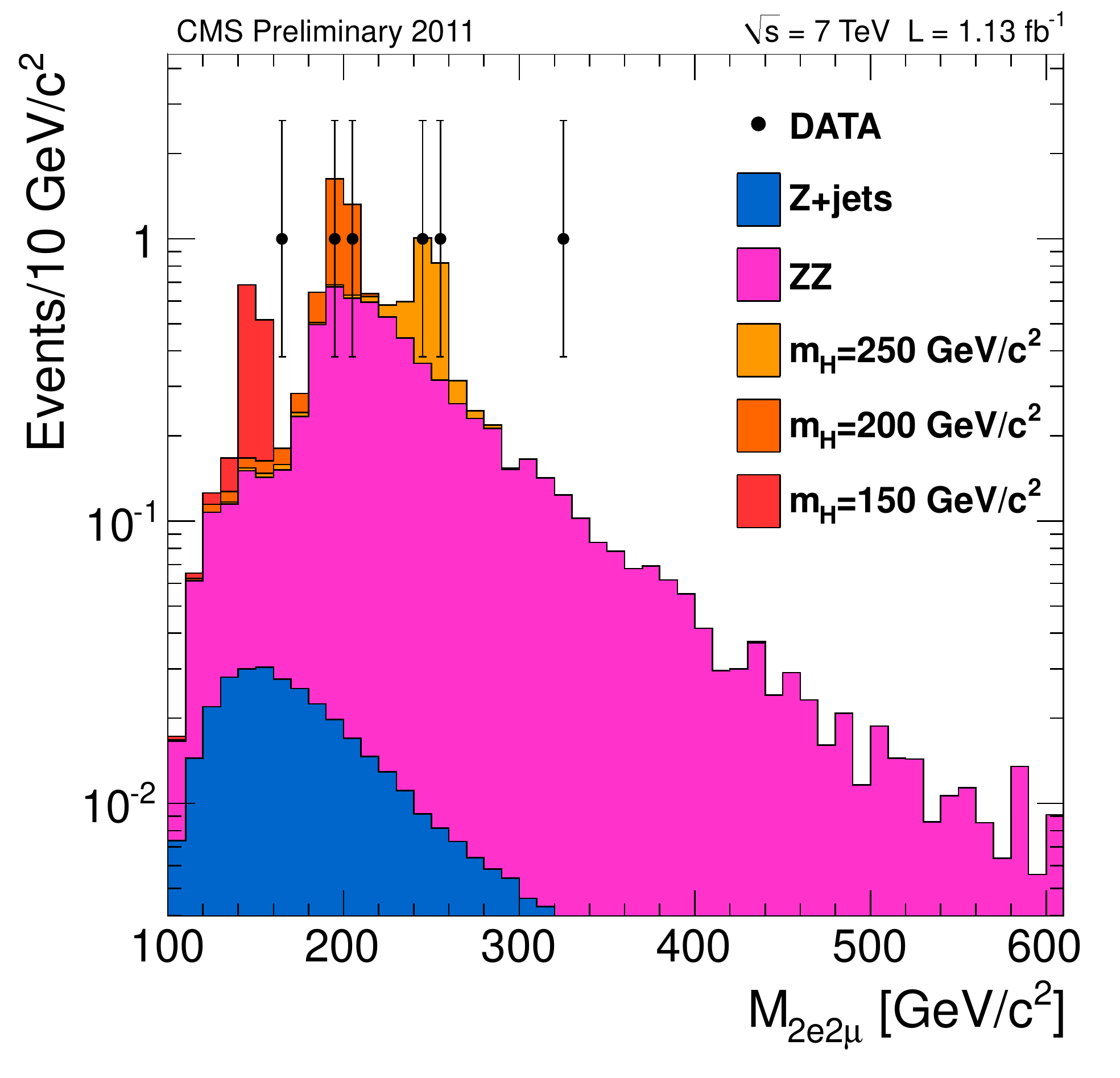}} 
\subfigure[]{\includegraphics[width=0.5\textwidth]{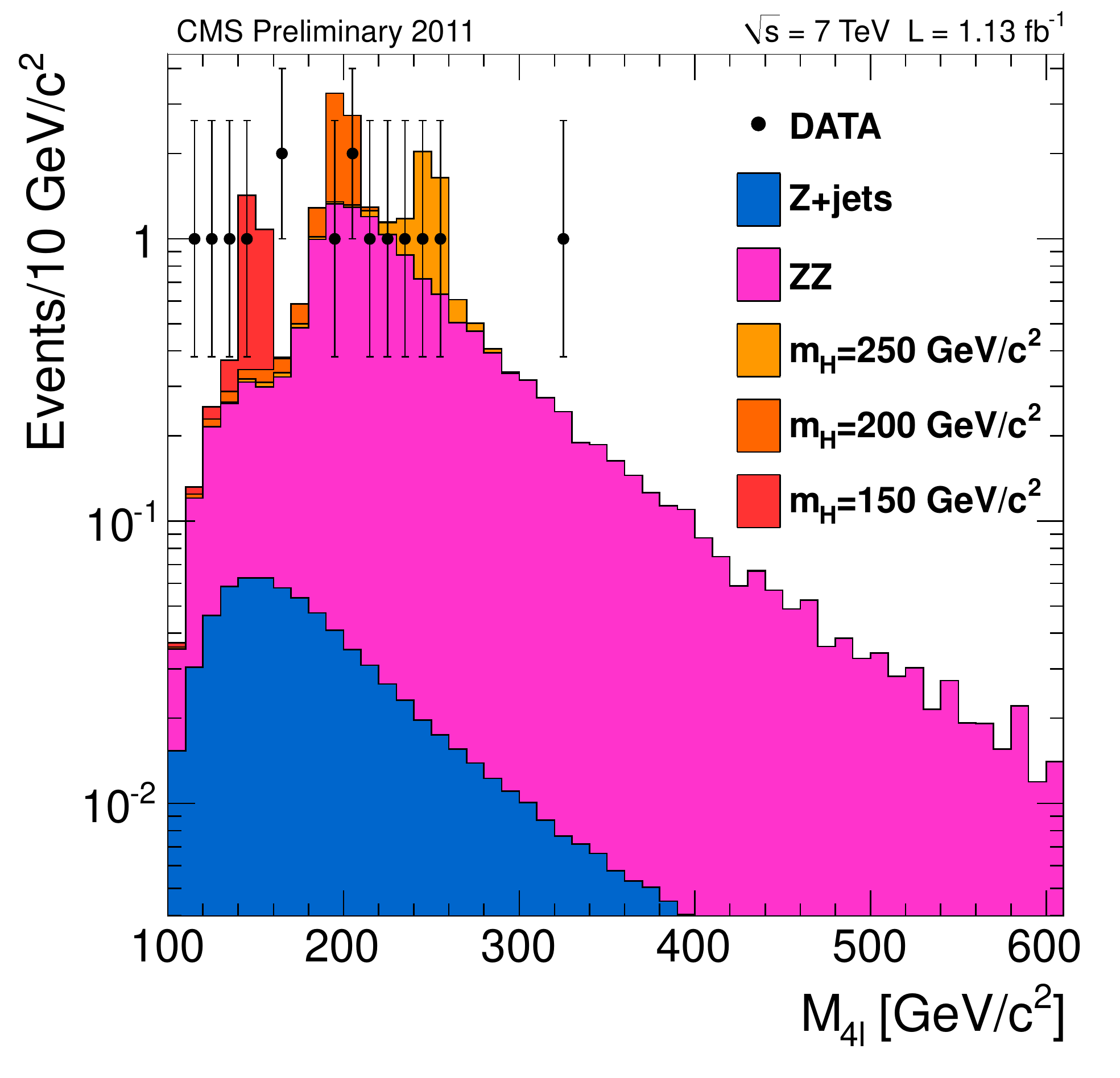}} \\
\end{tabular}
\caption{ Distribution of the four-lepton reconstructed mass for the baseline selection 
                in the (a) $4e$, (b) $4\mu$,  (c) $2e2\mu$, and (d) the sum of the $4\ell$ channels.
                Points represent the data, shaded histograms represent the signal and background expectations.
                The samples correspond to an integrated luminosity of ${\cal L} = 1.13 \, \ifb$.}
\label{fig:Mass4lBaseline}
\end{center}
\end{figure}

The reducible and instrumental backgrounds are very small or negligible. 
The number of events observed, as well as the background rates in the signal region within 
a mass range from $m_1 = 100~\GeVcc$ to $m_2 = 600~\GeVcc$, are reported for each final state
in Table~\ref{tab:FinalYields} for the baseline and high-mass selections.

\begin{table}[htb]
\begin{center}
\caption{Number of events observed, background and signal rates for each final state
in a mass range from $m_1 = 100~\GeVcc$ to $m_2 = 600~\GeVcc$
both for the baseline and high-mass selections.
For ZZ, Z+jets, $t\bar{t}$ and Z$b\bar{b}/c\bar{c}$ the data driven estimations are used,
for WZ the Monte Carlo estimation is used.
}
\label{tab:FinalYields}
\begin{tabular}{c|c|c|c}
\hline 
& \multicolumn{3}{l}{\hspace*{2.5cm} baseline } \\
& $4e$ & $4\mu$ & $2e2\mu$  \\
\hline 
ZZ &  2.76 $\pm$ 0.18 &  4.10 $\pm$ 0.27 & 6.72 $\pm$ 0.45                 \\ 
Z+jet & 0.37 $\pm$ 0.07  &  0.06 $\pm$ 0.01 &  0.39 $\pm$ 0.07             \\ 
${\rm Z}b\bar{b}/c\bar{c}, t\bar{t}$ & 0.01 $\pm$ 0.02 & 0.01 $\pm$ 0.01 & 0.02 $\pm$ 0.02 \\ 
WZ                   & $0.006\pm0.006$ & $0.006\pm0.006$ & $0.024\pm0.012$ \\ 
\hline                                           
All background       & $3.13\pm0.19$ & $4.17\pm0.27$ & $7.14\pm0.46$       \\ %
\hline 
m${_H} = 150 \GeVcc$ & $0.368\pm0.007$ & $0.637\pm0.009$ & $0.996\pm0.011$ \\ 
m${_H} = 200 \GeVcc$ & $0.816\pm0.011$ & $1.161\pm0.014$ & $1.907\pm0.018$ \\ 
m${_H} = 250 \GeVcc$ & $0.656\pm0.009$ & $0.885\pm0.010$ & $1.533\pm0.013$ \\ 
\hline
Observed & 3 & 6  & 6 \\ 
\hline
\multicolumn{4}{l}{} \\
\hline
& \multicolumn{3}{l}{\hspace*{2.5cm} high-mass} \\
& $4e$ & $4\mu$ & $2e2\mu$   \\
\hline 
ZZ & 2.50 $\pm$ 0.17 & 3.55 $\pm$ 0.23  &  6.10 $\pm$ 0.40                 \\ 
Z+jet &  0.14$\pm$ 0.06 & 0.004 $\pm$ 0.004 &   0.15 $\pm$ 0.06            \\ 
${\rm Z}b\bar{b}/c\bar{c}, t\bar{t}$  & - & - & -                          \\ 
WZ                   & - & - & $0.006\pm0.006$                             \\ 
\hline
All background       & $2.64\pm0.18$ & $3.55\pm0.23$ & $6.25\pm0.40$       \\ %
\hline 
m${_H} = 150 \GeVcc$ & $0.018\pm0.002$ & $0.030\pm0.002$ & $0.045\pm0.002$ \\ 
m${_H} = 200 \GeVcc$ & $0.765\pm0.011$ & $1.080\pm0.013$ & $1.801\pm0.017$ \\ 
m${_H} = 250 \GeVcc$ & $0.621\pm0.009$ & $0.833\pm0.010$ & $1.442\pm0.013$ \\ 
\hline
Observed & 0 & 2 & 6  \\ 
\hline
\end{tabular}
\end{center}
\end{table}

The measured distribution is seen to be compatible with the expectation from SM continuum 
production of ZZ$^{(*)}$ pairs.
We observe $N_{\rm obs}^{\rm baseline}= 15$ events for the baseline selection, in good agreement
with the expectation of $14.4\pm 0.6$ events from SM background evaluation. 
Six of the events are below the kinematic threshold of two on-shell Z's ($m_{\rm H} < 180 \GeVcc$),
while $1.9\pm 0.1$ background events are expected. 
The probability that the background fluctuates to the observed number of events is 1.3\%.
The events are not clustered, excluding the interpretation as the standard model Higgs boson. 
However we note that the six events form three mass pairs: two are close to each of the following masses 
$122$, $142$ and $165 \GeVcc$, respectively.
We observe $N_{\rm obs}^{\rm highmass}= 8$ events for the high-mass selection compared to an expectation of 
$12.4 \pm 0.5$ events from SM background evaluations.
This corresponds to a 13\% probability for the background to fluctuate to a value $\le N_{\rm obs}^{\rm highmass}$.

No attempt to recover electromagnetic final state radiation prior to imposing the analysis selection criteria is made. 
However, a posteriori all selected events were inspected for the presence of direct photons that might originate 
by inner bremsstrahlung from leptons in the final state.

The high-mass event selection and analysis which imposes the presence of two lepton pairs with invariant
masses in the range $60 < m_{\ell^+\ell^-} < 120 \GeVcc$ is used to provide a measurement of the
total cross section $ \sigma (pp \rightarrow {\rm ZZ} + X) \times {\cal B}({\rm ZZ} \rightarrow 4\ell) $.
This measured cross section is obtained via:
\begin{equation}\label{eq:ZZresonance}
\sigma (pp \rightarrow {\rm ZZ} + X) \times {\cal B}({\rm ZZ} \rightarrow 4\ell)
= \frac{\sum N_{\rm obs}(i_{\rm ch}) - N_{\rm{back}}(i_{\rm ch})}
{{\rm{A_{4\ell}}} \times {\epsilon_{\rm{ZZ} \rightarrow 4\ell}} \times {\cal L}}
\end{equation}
Where $i_{\rm ch}$ means 4e, $4\mu$, 2e2$\mu$; $\rm{A}$ is the acceptance of the detector,
the efficiency $\epsilon_{\rm{ZZ} \rightarrow 4\ell}$ is the ratio between
the $ZZ$ events that survive the selection cuts  and the generated $ZZ$ events within the acceptance of the detector.
The fiducial and kinematic acceptance is defined by the fraction of events with four
final state leptons satisfying $p_{T} > 7 (e), 5(\mu)$ within $| \eta^e | < 2.5$ and $| \eta^{\mu} | < 2.4$.
The generated $ZZ$ events are the fraction of the PYTHIA MC sample used in the
analysis with $60 < m_{\rm Z} < 120 \GeVcc$.

The total cross section for a pair of Z bosons in the mass range
$60 < m_{\rm Z} < 120 \GeVcc$ is found to be
$$ \sigma (pp \rightarrow {\rm ZZ} + X) \times {\cal B}({\rm ZZ} \rightarrow 4\ell)
       = {\rm 17.5  ^{+6.3 } _{-4.4 }  (stat.) \pm 0.7 (syst.) \pm 1.1 (lumi.) \, fb} \,\, .$$
The measured cross section agrees within about 1.5 standard deviations with the expectation from the
SM \cite{MCFM}.

\begin{figure}[htb]
 \begin{center}
 \subfigure[]{\includegraphics[width=0.49\textwidth]{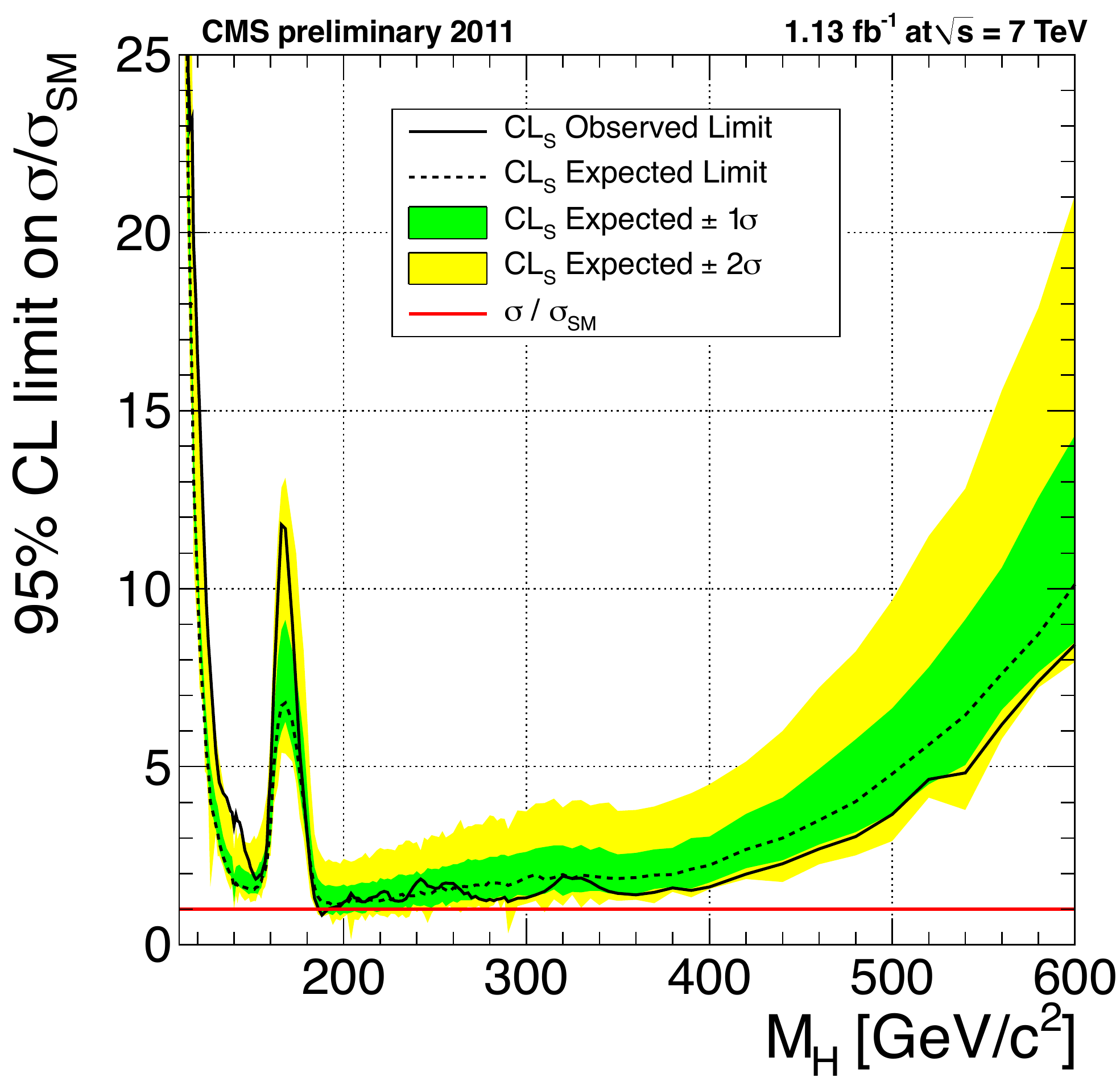}}
 \subfigure[]{\includegraphics[width=0.49\textwidth]{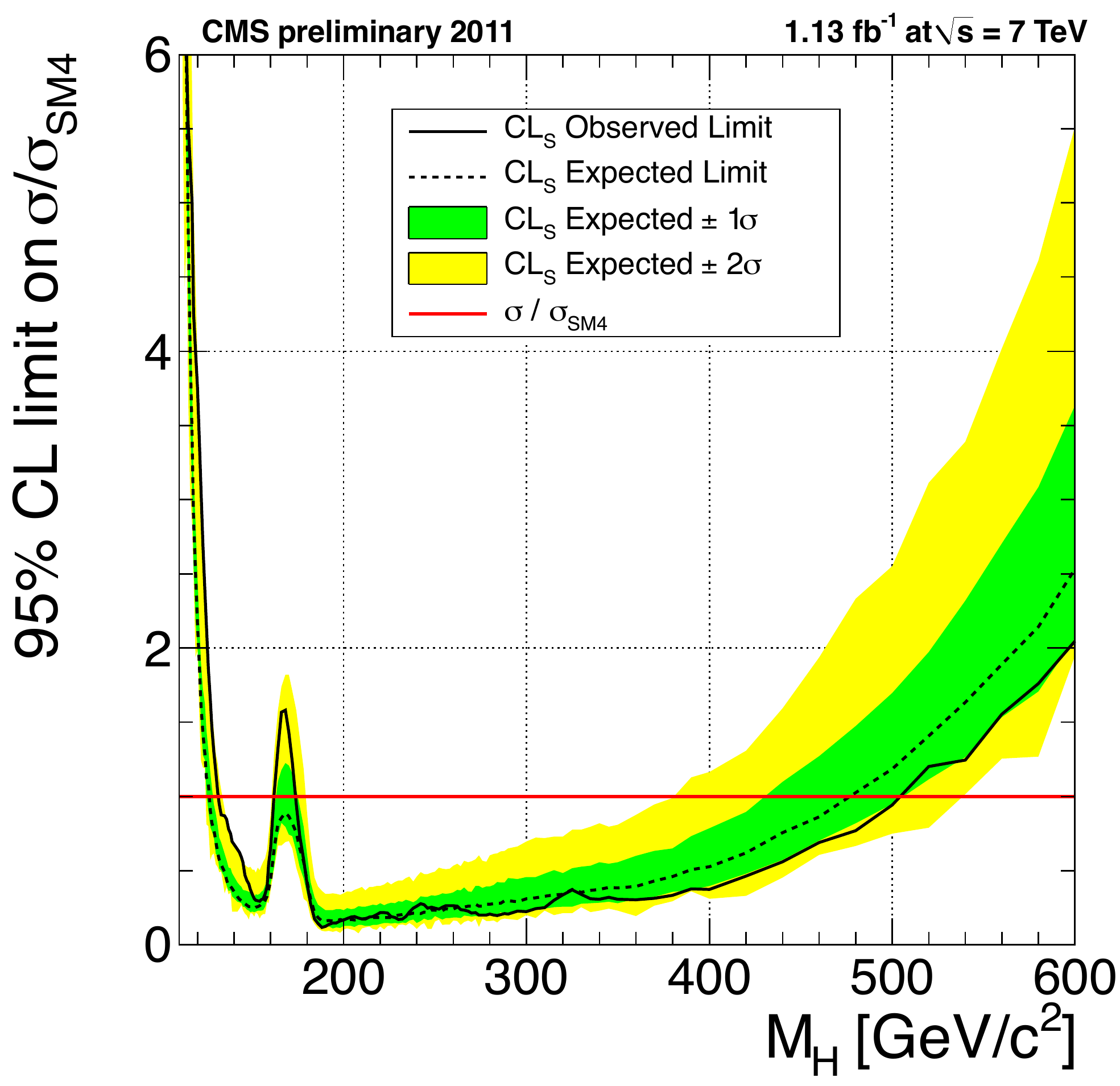}}
   \caption{
    The mean expected and the observed upper limits at 95\% C.L. 
    on $\sigma (pp \rightarrow {\rm H} + X) \times {\cal B}({\rm ZZ} \rightarrow 4\ell)$ 
    for a Higgs boson in the mass range 120-600~$\GeVcc$, for an integrated luminosity of 
    1.13$ \, \ifb$ using the CL$_{\rm s}$ approach.
    The expected ratios for (a) the SM and for (b) the SM with a fourth-fermion family (SM4) are presented.
    The results are obtained using a shape analysis method.}
   \label{fig:xsLimCutandCountDATA}
 \end{center}
\end{figure}

The exclusion limits for a SM-like Higgs Boson are computed for a large number of mass points
in the mass range 110-600~$\GeVcc$, and using the predicted signal and background shapes.
The choice of the spacing between Higgs mass hypotheses is driven by either detector resolution, 
or the natural width of the resonance, depending on which dominates. 
The signal shape is determined using 17 simulated samples covering the full mass range.  
The shapes for each simulated sample are fit using a function ($f^{\rm MC}$) obtained as
a convolution between a Breit-Wigner-like probability density function to describe the theoretical 
resonance line shape, and a Crystal-Ball function to describe the detector effects.  
The parameters of the Crystal-Ball function are interpolated for the Higgs boson mass points 
where there is no simulated sample available.
The mass shapes for the backgrounds  are determined by fits to the MC samples,
with full detector simulation, while the normalization 
is taken from the overall event yield estimates as described in previous sections.

As a cross check the cut-and-count method has been used. To calculate the exclusion limits, the shapes are integrated 
in a given mass window to extract the signal and background yields.
The mass window is chosen for each mass point to find the optimal expected limit, computed with a Bayesian 
approach with flat prior. 
The resulting mass windows are asymmetric and their width accounts for the mass migrations from generated 
to reconstructed quantities, the mass resolution and the intrinsic width for a given central mass hypothesis.
Similar exclusion limits results are obtained for the cut-and-count method.

The observed and mean expected 95\% CL upper limits on Higgs 
$\sigma (pp \rightarrow {\rm H} + X) \times {\cal B}({\rm ZZ} \rightarrow 4\ell)$
in a shape analysis method, obtained for Higgs masses in the range 110-600~$\GeVcc$
are shown in Fig.~\ref{fig:xsLimCutandCountDATA}. 
The limits are made using a CL$_{\rm s}$ approach,  
for the expected ratios to the SM, and in the context of a SM extension by a sequential 
fourth family of fermions with very high masses
(SM4)~\cite{Li:2010fu,Schmidt:2009kk,FourthGen}.
The bands represent the $1\sigma$ and $2\sigma$ probability intervals around the expected limit.
The expected background yield is small hence the $1\sigma$ 
range of expected outcomes includes pseudo-experiments with
zero observed events.
The lower edge of the $1\sigma$ band therefore corresponds already to the most
stringent limit on the signal cross section, as fluctuations below that value are not possible.
We account for systematic uncertainties in the form of nuisance parameters with a log-normal 
probability density function.
The exclusion limits extend at high mass beyond the sensitivity of previous collider experiments.
The expected limits reflect the dependence of the branching ratio ${\cal B}({\rm H} \rightarrow {\rm ZZ})$
on the Higgs boson mass. The worsening limits at high masses arise from the decreasing signal cross section. 
The reduced sensitivity around $m_{\rm H} = 160 \GeVcc$ and for the low masses arise from the very small 
${\rm H} \rightarrow {\rm ZZ}$ branching ratio in these regions.

In the current study the exclusion limit is compared to the cross section
for on-shell Higgs production and decay in the zero-width approximation, 
and acceptance estimates are obtained with Monte Carlo simulations
that are based on ad-hoc Breit-Wigner distributions for describing the
Higgs-boson propagation.

\section{Conclusions}\label{sec:Conclusions}

The results of a search for a standard model Higgs boson produced
in pp collisions at $\sqrt{s} = 7 \TeV$ and decaying in ${\rm ZZ}^{(*)}$ have been 
presented for the first time by the CMS Collaboration
in the leptonic Z decay channel ${\rm ZZ}^{(*)} \rightarrow 4\ell$, with $\ell = e,\mu$. 
Simple sequential sets of lepton reconstruction, identification and isolation cuts and a set of kinematic 
cuts have been introduced to define a common baseline for the search at any 
Higgs boson mass $m_{\rm H}$ in the range $ 100 < m_{\rm H} < 600 \GeVcc$. 
The instrumental background from Z+jets and the reducible backgrounds from
Z$\rm{{b}}\rm{{\bar{b}}}$ and $\rm{{t}}\rm{{\bar{t}}}$, with misidentified primary leptons,
are shown to be negligible over most of the mass, with a small contamination remaining 
at low mass.

Fifteen events are observed in the $2e2\mu$, $4e$ and $4\mu$ channels for
an integrated luminosity of  $1.13 \pm 0.07 \; \ifb$, while $14.4 \pm 0.6$ events are expected 
from standard model background processes. 
The distribution of events is compatible with the expectation from the standard model  continuum 
production of Z boson pairs from  $q\bar{q}$ annihilation and $gg$ fusion. 
No clustering of events is observed in the measured $m_{4\ell}$ mass spectrum.
Six of the events are below the kinematic threshold of two on-shell Z's ($m_{\rm H} < 180 \GeVcc$),
while $1.9\pm 0.1$ background events are expected.
The probability that the background fluctuates to the observed number if events is 1.3\%.
Using the high-mass selection which contains eight events, a total cross section 
for a pair of Z bosons in the mass range $60 < m_{\rm Z} < 120 \GeVcc$ has been measured 
to be in agreement with the predicted value.
Upper limits obtained at 95\% CL on the cross section$\times$branching ratio for a Higgs boson with 
standard model-like decays exclude cross sections from about one to two times the expected 
standard model cross section for masses in the range $180 < m_{\rm H} < 420 \GeVcc$. 
Upper limits obtained in the context the standard model with a fourth fermion family, 
exclude a Higgs boson with a mass in the ranges 138-162  $\GeVcc$  or 178-502 $\GeVcc$ 
at 95\% CL.

\bigskip 

\bibliographystyle{plain}
\bibliography{bibliography}

\end{document}